\documentclass{article}
\usepackage{spconf}

\usepackage{amsfonts,amssymb,amsmath,amsthm}
\usepackage{subfigure}
\usepackage{graphicx,psfrag}
\usepackage[footnotesize]{caption}

\usepackage{bbm}
\usepackage{bbold}
\usepackage{float}
\usepackage{subfig}
\usepackage{algorithm2e}
\usepackage{color}
\usepackage{multirow}
\usepackage{cite}
\usepackage{enumitem}

\input alphabet
\input jidef

\def\dd{\,d}

\def\eqname{Eq.~}
\def\eqnames{Eqs~}
\def\tabname{Table~}

\def\figname{Figure~}

\def\chapname{Chapter~}

\def\parname{\S~}
\def\parnames{\S~}
\def\sectname{Section~}

\def\<{``}\def\>{''}

\def\GIG{\ensuremath{\mathit{GIG}}}
\def\GH{\ensuremath{\mathit{GH}}}

\newcommand{\N}{\mathcal{N}}

\renewcommand{\thepage}{}

\usepackage[top=.75in,left=0.63in, right=0.63in, bottom=.75in]{geometry}

\title{A partially collapsed sampler for unsupervised nonnegative spike train restoration}
\name{M.C. Amrouche$^1$, H. Carfantan$^1$ and J. Idier$^2$}
\address{$^1$Institut de Recherche en Astrophysique et Planetologie,\\ Université de Toulouse, CNRS/UPS/CNES, Toulouse, France\\
         $^2$Laboratoire des Sciences du Numérique de Nantes, CNRS/ECN, Nantes, France\\
         mamrouche@irap.omp.eu, hcarfantan@irap.omp.eu, jerome.idier@ls2n.fr
         }
\date{\empty} 

\renewenvironment{abstract}{\bf\small {\em\ Abstract---}}{}

\begin{document}

\maketitle

\begin{abstract} In this paper the problem of restoration of non-negative sparse signals is addressed in the Bayesian framework. We introduce a new probabilistic hierarchical prior, based on the Generalized Hyperbolic (GH) distribution, which explicitly accounts for sparsity. This new prior allows on the one hand, to take into account the non-negativity. And on the other hand, thanks to the decomposition of GH distributions as continuous Gaussian mean-variance mixture, allows us to propose a partially collapsed Gibbs sampler (PCGS), which is shown to be more efficient in terms of convergence time than the classical Gibbs sampler.
\end{abstract}

\section{Introduction}
\label{sec:introduction}
This paper tackles the restoration of a sparse non-negative signal, observed though a linear operator and corrupted by additive white noise. The $N\times 1$ observed signal $\yb$ can be written as $\yb = \Hv \zb +\epsilonb$, where $\zb$ is the $K\times 1$ sparse non-negative signal, $\Hv$ is a $N\times K$ matrix and $\epsilonb$ models the perturbations. 

In the literature, sparse signal restoration problems arise in different fields such as reflection seismology, astronomy and compressed sensing. The objective is to find a sparse representation of a given signal $\yb$ that is a linear combination of a limited number of elements (atoms) taken from a given dictionary $\Hv$. This problem is often referred to as subset selection because it consists in selecting a subset of columns of $\Hv$. Mathematically, this can be formulated as the minimization of the squared error $\norm{\yb - \Hv \zb}^2$ subject to $\norm{\zb}_0 < S$, where $\norm{\cdot}$ and $\norm{\cdot}_0$ respectively stands for the Euclidean norm, and the $\ell_0$ pseudo-norm. 
This yields a combinatorial discrete problem known to be NP-hard \cite{natarajan95}.

One alternative is the \textit{convex relaxation} of the problem which substitutes the $\ell_0$ pseudo-norm by the $\ell_1$ norm \cite{chen,donoho08},
the sparsity of the solutions coming from the non smooth character of the $\ell_1$ norm at zero. 
\textit{Greedy algorithms}, such as \textit{Matching Pursuit} and its improved versions \textit{Orthogonal Matching Pursuit} and \textit{Orthogonal Least Squares}, form another class of methods. The main idea is to iteratively enlarge or reduce by one the set of active atoms.

Non-negative adaptations have been proposed for both \textit{convex relaxation} and \textit{greedy algorithms}. For example, the $\ell_1$ relaxation can be easily extended to the non-negative setting \cite{barbu16,efron}. Non-negative extensions of greedy algorithms have been also introduced \cite{bruckstein08,yaghoobi15,Nguyen19}.



On the other hand, a hierarchical Bayesian model which explicitly accounts for sparsity has been proposed for spike deconvolution, namely the Bernoulli-Gaussian (BG) model \cite{kormylo82,champagnat96}. Deterministic optimization algorithms \cite{champagnat96} and Markov chain Monte Carlo techniques (MCMC) \cite{cheng96} are used to compute, respectively, the Maximum a Posteriori (MAP) and the Posterior Mean (PM) estimators. In the sparse train deconvolution context, the aforementioned methods have no theoretical guarantee regarding the exact recovery of the signal. However, the PM estimator has been shown empirically to give better results than greedy algorithms and convex relaxation \cite{bourguignon11}. Moreover, the Bayesian framework allows one to estimate the model hyper-parameters (unsupervised case). 

In the MCMC framework, a non-negative adaptation of the BG model was studied in \cite{mazet05}, based on a Bernoulli-Truncated-Gaussian (BTG) model. Vector $\zb$ is modelled as a couple of variables $(\qb,\xb)$ \emph{i.e.}, independent, identically distributed (i.i.d.) Bernoulli variables $\qb_k$, i.i.d. centered truncated Gaussian amplitudes $\xb_k$ when $\qb_k=1$, and $\xb_k=0$ when $\qb_k=0$. Posterior Mean estimation of $\qb$ and $\xb$ is then obtained using an MCMC method. The latter yields satisfactory results, but it requires a high computational cost.
In \cite{vandyk08}, some strategies and tools are presented to construct a partially collapsed Gibbs sampler (PCGS) with known stationary distribution and faster convergence. The partially collapsed Gibbs sampler substitutes some of the posterior conditional distributions with marginal distributions of the joint posterior. Such a technique has proven its efficiency for BG deconvolution \cite{boudineau16,ge11} where the amplitudes $\xb$ can be marginalized out from the joint posterior thanks to their \textit{Gaussianity}. Unfortunately, partially collapsed sampling cannot be as easily adapted to the context of non-negative BG restoration.

The aim of this paper is to implement partially collapsed sampling in the context of non-negative sparse restoration. Our main contribution is to propose a new prior based on the \textit{Generalized Hyperbolic} (GH) distribution. On the one hand, the proposed prior mimics the BTG prior to take into account the non-negativity. On the other hand, the decomposition of GH distributions as continuous Gaussian mean-variance mixture allows us to  marginalize the amplitudes, enabling the construction of a partially collapsed Gibbs sampler which is shown hereafter to converge more rapidly than the BTG version.
The organization of this paper is as follows. An overview of the Generalized Hyperbolic distribution and its properties are given in Section~\ref{sc:GHdis}. Section~\ref{sc:bayesian} introduces the Bernoulli-Generalized-Hyperbolic (BGH) model in the Bayesian framework. Its corresponding partially collapsed Gibbs sampler is presented in Section~\ref{sc:sampler}. Simulation results are given in Section~\ref{sc:simulation} to compare the BGH and the BTG samplers both in terms of signal restoration and convergence. Finally, conclusions are drawn in Section~\ref{sc:conclusion}.

\section{Generalized hyperbolic distributions} \label{sc:GHdis}
\subsection{Description and properties}
Generalized Hyperbolic (GH) laws form a five-parameter family $(\lambda,\alpha,\beta,\delta,\mu)$ introduced by Barndorff-Nilsen in \cite{barndorff-nielsen77a}. If a random variable $X \sim \GH(\lambda,\alpha,\beta,\delta,\mu)$ then its probability density is given by
\begin{equation*}\label{eq:GH}
p_{X}(x) = \frac{(\gamma/\delta)^\lambda}{\sqrt{2\pi}K_\lambda (\delta\gamma)}\frac{K_{\lambda -\frac1{2}}(\alpha\sqrt{\delta^2 + (x-\mu)^2})}{(\sqrt{\delta^2 + (x-\mu)^2}/\alpha)^{\frac1{2}-\lambda}} e^{\beta(x-\mu)},
\end{equation*}
where $\gamma^2 = \alpha^2 - \beta^2$ and $K_\lambda$ is the modified Bessel function of third kind.
We will take advantage of two interesting properties of the GH distributions:
\begin{itemize}[leftmargin=*]
\item GH distributions have the property of being invariant under affine transformations. If  $X \sim \GH(\lambda,\alpha,\beta,\delta,\mu)$, then $Y = aX + b \sim \GH(\lambda,\alpha/a,\beta/a,a\delta,a\mu+b)$.
\item They can be expressed as continuous normal mean-variance mixtures
\end{itemize}
$$
p_{X}(x) = \int_0^{\infty} p_{X|W}(x|w)\,p_{W}(w)\dd w
$$
where $X|W \sim \mathcal{N}(\mu+\beta W,W)$ and each random variance $W$ follows a \textit{Generalised Inverse Gaussian} \cite{joergensen82} distribution $\GIG(\lambda,\gamma,\delta)$ with the probability density	
\begin{equation*}
p_{W}(w)=\frac{(\gamma/\delta)^\lambda}{2K_\lambda (\delta\gamma)} w^{\lambda - 1} \exp\pth{-\fracp{1}{2}\pth{\delta^2 w^{-1} + \gamma^2 w}}, ~w>0.
\end{equation*}
Therefore, a GH random variable can be seen as a hierarchical model: in order to generate a GH sample, a random variance $W\sim\GIG(\lambda,\gamma,\delta),$ is first drawn, and then $X|W$ with a normal distribution $\mathcal{N}(\mu+\beta W,W)$.


\subsection{Truncated Gaussian approximation}
One easy way to estimate the parameters of the GH distribution that best approximate a given distribution, is to first generate $N$ samples $\sb = \{\sb_1,\cdots,\sb_N\}$ according to it, and then to estimate the GH parameters using the Maximum-likelihood estimator \cite{breymann13}. As $N \to \infty$, this is equivalent to minimizing the Kullback-Leibler divergence between the two distributions.

Let $\nub_\N=(\lambda_\N,\alpha_\N,\beta_\N,\delta_\N,\mu_\N)$ denote the parameters that best approximate the truncated version of the normalized Gaussian $\N^{+}(0,1)$ with the GH distribution:
\begin{equation}
X \sim \GH(\nub_\N) \approx \mathcal{N}^+(0,1),
\end{equation}
Since the GH distribution is closed under affine transformations, we can approximate the truncated version of any other centered Gaussian $\N^{+}(0,\sigma_x^2)$ as follows: 
\begin{equation}
\GH(\lambda_\N,\alpha_\N / \sigma_x,\beta_\N / \sigma_x,\delta_\N \sigma_x,\mu_\N\sigma_x) \approx \mathcal{N}^+(0,\sigma_x^2).
\end{equation}
As the parameters of this distribution are fixed given $\sigma_x$, we will write $\GH_\N(\sigma_x^2)$ for sake of simplicity. Similarly, we write $\GIG_\N(\sigma_x^2)$ instead of $\GIG(\lambda_\N,\gamma_\N/\sigma_x,\delta_\N \sigma_x)$.
%

\section{Bayesian framework}
\label{sc:bayesian}
\subsection{Bernoulli-Generalized-Hyperbolic prior}
In the following we will consider that $\zb = (\qb,\xb)$ where $\qb_k \sim\mathcal{B}(\lambda),\forall k=1,\cdots,K	$ and:
\begin{equation*}\label{eq:BGHModel}
\begin{split}
\xb_k | \qb_k = 1 &\sim   
	\left\{\begin{tabular}{l}
	$\wb_k \sim \GIG_\N(\sigma_x^2)$ \\
	$\xb_k|\wb_k \sim \N(\sigma\mu_\N + \wb_k\beta_\N/\sigma,\wb_k),$ 
	\end{tabular}\right. \\
\xb_k | \qb_k = 0 &\sim \delta(\xb_k), \\
\end{split}
\end{equation*}
leading to the Bernoulli-Generalized-Hyperbolic prior. 
\subsection{Posterior distribution}
We consider an independent noise sequence $\{\epsilonb_k\}_k$, zero-mean, Gaussian, with variance $\sigma^2$. The posterior distribution can be written as follows:
\begin{equation}
p(\qb,{\xb},\wb|\yb) \propto \exp\left(-\fracp{1}{2\sigma^2} \norm{ \yb - \overline{\Hv}\overline{\xb} }^2\right) p(\qb,{\xb},\wb)
\end{equation}
where $\overline{\xb}$ gathers the elements of the vector $\xb$ for which $\qb_k = 1$ and similarly $\overline{\Hv}$ concatenates the columns of the matrix $\Hv$ for which $\qb_k = 1$. 


For an unsupervised estimation, a prior distribution can be introduced for the hyper-parameters $\thetab = [\lambda,\sigma^2,\sigma_x^2]$ independent of the other parameters, and the posterior distribution can be written:
\begin{equation}\label{eq:posterior}
p(\qb, \xb, \wb, \thetab | \yb) \propto p(\qb, \xb, \wb| \yb, \thetab)p(\thetab)
\end{equation}

\subsection{Partially marginalized posterior distribution}
From \eqref{eq:posterior}, it is straightforward to deduce that $\overline{\xb}|\qb,\wb,\yb,\thetab$ follows a multivariate Gaussian
$\N(\etab,\Gammab)$, with
\begin{align*}
\etab &= \sigma^{-2}\Gammab \overline{\Hv}^T \yb + \Gammab{\Wv}^{-1} {\mub}_x,\\
\Gammab &= \bigpth{\sigma^{-2}\overline{\Hv}^T \overline{\Hv} + {\Wv}^{-1}}^{-1},
\end{align*}
where ${\mub}_x = \sigma\mu_\N\mathbb{1}_L + \frac{\beta_\N}{\sigma} {\wb}$ and $L = \sum_k \qb_k$. As a consequence, one can easily calculate the marginalized posterior distribution with respect to $\xb$
\begin{align*}
p(\qb,\wb,\thetab|\yb) &\propto  |\Bv|^{-1/2} ~ P(\qb|\lambda) p(\thetab) \prod_{k} p(\wb_k) \\ 
\times &\exp\pth{-\fracp{1}{2}( \yb^T \Bv^{-1} \yb + \mub_x^T \Cv^{-1} \mub_x  -2\yb^T \Dv \mub_x)} \\
\end{align*}
with $\Wv = \diag\{\wb\}$ and:
\begin{equation*}
\begin{array}{lll|c|}
\Bv^{-1} &=& \sigma^{-2}\Iv - \sigma^{-4} \overline{\Hv} \Gammab \overline{\Hv}^T  & (N\times N) \\
\Cv^{-1} &=& \Wv^{-1} - \Wv^{-1}\Gammab \Wv^{-1} & (L\times L) \\
\Dv &=& \sigma^{-2} \overline{\Hv} \Gammab \Wv^{-1} & (N\times L)
\end{array}
\end{equation*}

\section{Partially Collapsed Gibbs Sampler}\label{sc:sampler}
\label{sec:second-section}
While sampling $\xb|\qb,\wb$ is an easy task ($\overline{\xb}$ is Gaussian), 
sampling $\qb$ and $\wb$ need to be considered  in a special framework as the variable $\wb_k$ is defined only if $\qb_k = 1$ which implies that the posterior distribution is defined in a space with a varying dimension. 
A general framework for Metropolis-Hastings algorithms was introduced in \cite{green}, namely the Reversible-Jump (RJ) MCMC methods, in order to manage jumps between subspaces of different dimensions in stochastic sampling algorithms.
The sampling strategy adopted here is as follows (Algorithm~\ref{alg:Sampler}). At each iteration $i$, first sample $\qb^{(i)}$ and $\wb^{(i)}$ using the RJ-MCMC framework as shown in Algorithm~\ref{alg:SamplerRJ} (see \ref{sc:RJ} for explanations), then sample $\xb^{(i)}|\qb^{(i)},\wb^{(i)}$, and finally sample the hyper-parameters $\thetab$ according to their posterior.

\RestyleAlgo{boxruled}
\begin{algorithm}[h]
At each iteration $i$:
\setlist{nolistsep}
\begin{enumerate}[leftmargin=*]
	\item for all $k$ in $1,\cdots,K$:
	\begin{enumerate}
		\item draw $\qb_k,\wb_k|\qb_{-k},\wb_{-k},\thetab,\yb$ according to 
		
		Algorithm~\ref{alg:SamplerRJ}
	\end{enumerate}
	\item draw $\xb|\qb,\wb,\thetab,\yb$ (Gaussian distribution)
	\item draw $\thetab|\qb,\wb,\xb,\yb$ according to their conditional 
	
	posterior distributions
\end{enumerate}
 \caption{Partially collapsed Gibbs sampler}\label{alg:Sampler}
\end{algorithm}

\subsection{Reversible-Jump step} \label{sc:RJ}
For the problem considered here, two states can be distinguished; whether $\qb_k = 1$ and $\wb_k$ is defined, or $\qb_k = 0$ and $\wb_k$ is not defined. The RJ-MCMC framework allows to jump between these two states using the moves hereafter:
\begin{enumerate}[noitemsep]
\item \textit{Birth}: from $\qb_k = 0$ propose $\qb_k' = 1,\wb_k'$,
\item \textit{Death}: from $\qb_k = 1,\wb_k$ propose $\qb_k' = 0$,
\item \textit{Update}: from $\qb_k = 1,\wb_k$ propose $\qb_k' = 1,\wb_k'$.
\end{enumerate}
note that the move from $\qb_k = 0$ to $\qb_k' = 0$ is not of interest as this proposal is always accepted, and cannot introduce any change. 

In the following, we denote by $p_{uu'}$ the probability of proposing a move from the state $u$ to $u'$. 
Since we have chosen to systematically propose a \textit{birth} move when $\qb_k = 0$ then $p_{01} = 1$. Otherwise, when $\qb_k = 1$, it is reasonable to randomly propose either a \textit{death} or an \textit{update} move with equal probability $p_{10} = p_{11} = \frac{1}{2}$.

The $\wb_k'$ candidates are proposed according to the following proposal distributions. When a \textit{birth} move is chosen, it seems natural to propose $\wb_k'$ according to its prior distribution $q_{01}(\wb_k')\sim \GIG_\N(\sigma_x^2)$. Also, if a \textit{death} is selected the variable $\wb_k'$ disappears meaning that the proposal is deterministic, hereafter for notational convenience	we will write $q_{10}(\wb_k') = 1$ (for more details, one may refer to \cite[Remark 4.2]{waagepetersen01}) . Finally, for the \textit{update} move, a mix of two proposal distributions is considered: the first one is the prior distribution $q^{(1)}_{11}(\wb_k')\sim \GIG_\N(\sigma_x^2)$  allowing a better exploration of the feasible domain of $\wb_k$, while the second was empirically determined to maximize the acceptance probability, and thus to produce a better exploration of the posterior, according to:
$$
q^{(2)}_{11}(\wb_k') \sim
\GIG\left(\lambda_\N - \fracp{1}{2},\fracp{1}{\sigma_x}\sqrt{\gamma_\N^2 + \beta_\N^2},\sigma_x\sqrt{\delta_\N^2+ \mu_\N^2}\right)
$$
In practice, we have noticed that alternating between these two proposals with equal probabilities improves the mixing property of the Markov chain. The candidates are accepted according to the following probability:
\begin{equation}
\alpha_{uu'} = \min \left[1,\frac{p(\yb,\qb',\wb')\,p_{u'u}\,q_{u'u}(\wb_k')}{p(\yb,\qb,\wb)\,p_{uu'}\,q_{u'u}(\wb_k)} \right].
\label{eq:accept}
\end{equation}
This ensures the reversibility, and thus the invariance of the Markov chain with respect to the posterior distribution (see \cite{green,waagepetersen01}).
Direct evaluation of \eqref{eq:accept} is inefficient from the computational viewpoint. Therefore, we have adapted and extended the numerical implementation technique introduced in \cite{ge11} to reduce the computation and memory load. A key point is to recursively handle Cholesky factors instead of matrices $\Bv$, $\Cv$ and $\Dv$. Due to lack of space, calculation details cannot be given here.
%
%
\RestyleAlgo{boxruled}
\begin{algorithm}[t]
\SetAlgoLined
\eIf{$\qb_k$ = 0}{
--~propose a \textit{birth} move:

$\qb_k' = 1$ and $\wb_k' \sim q_{01}(\wb_k')$, then accept the proposition with probability $\alpha_{01}$.
}
{
--~with probability $p_{10}$ propose a \textit{death} move:

$\qb_k' = 0$ and $\wb_k' = \O,$ then accept the proposition with probability $\alpha_{10}$.

--~with probability $p_{11} = 1 - p_{10}$ propose an \textit{update} move:

$\qb_k' = 1$ and $\wb_k' \sim q^{(1)}_{11}(\wb_k')$ or $\wb_k' \sim q^{(2)}_{11}(\wb_k')$ with a probability $\frac12,$ then accept the proposition with probability $\alpha_{11}$.

}
 \caption{Sampling $\qb_k$ and $\wb_k$ using the RJ framework}\label{alg:SamplerRJ}
\end{algorithm}
\section{Simulation tests}\label{sc:simulation}
\begin{figure}[tbp]
\begin{tabular}{@{}l@{~}c@{}}
\rotatebox{90}{\hspace{.9cm}$h_n$}
&\includegraphics[width=.45\textwidth]{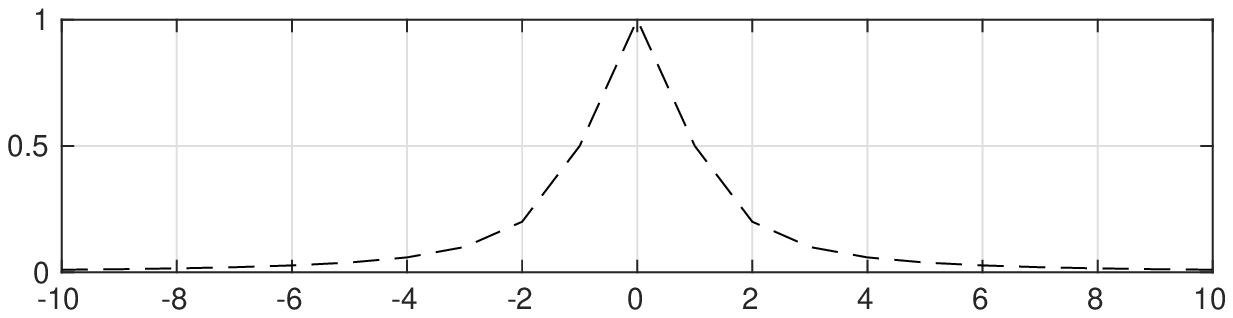} \\[-.1cm]
&$n$\\[-.1cm]
\rotatebox{90}{\hspace{1cm}$y_n$}
&\psfrag{Noiseless data}{\tiny Noiseless data}
\psfrag{Noisy data}{\tiny Noisy data}
\includegraphics[width=.45\textwidth]{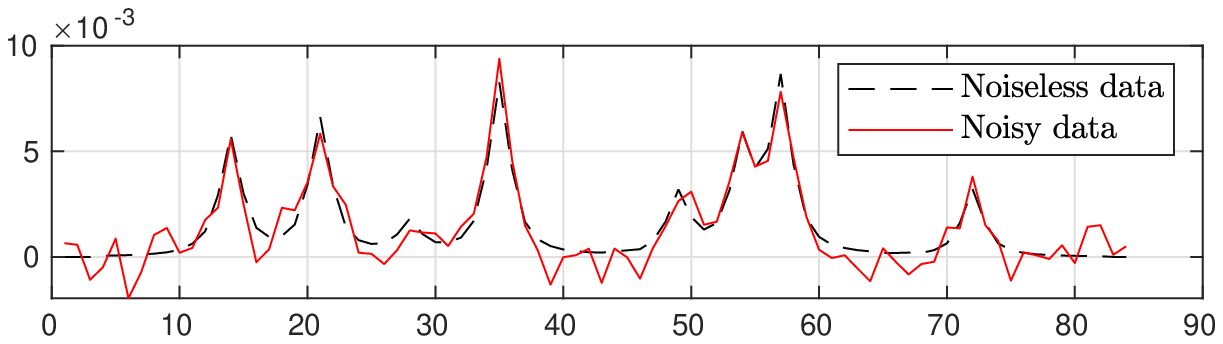}\\[-.1cm]
&$n$\\[-.2cm]
\end{tabular}
\caption{Impulse response $h_n$ and simulated data $y_n$ for the test scenario.}
\label{fig:Data}
\end{figure}

In this section, our goal is to compare the BTG and BGH samplers, both in terms of signal restoration and convergence rate. We have designed a test scenario similar to the one used in \cite{mazet05}. The observed signal $\yb$ (of length $N = 84$) is corrupted by Gaussian noise with $\sigma^2 = 5.5\times 10^{-7},$ corresponding to a signal-to-noise ratio (SNR) of $10$ dB (Fig~\ref{fig:Data}), and the impulse response was chosen to be of length $K=21$ and of the following form 
$
\hb_n = \fracp{s^2}{s^2+n^2}
$
where $s$ is an unknown scale parameter. Here, we propose to sample $s$ using a Metropolis-Hastings step. Furthermore, we consider an unsupervised situation where the hyper-parameters $\thetab = [\lambda,\sigma_x^2]^T$ are also unknown and sampled, while the noise variance is assumed to be known.

%
To empirically compare the convergence speed of the different samplers, we have used
Brooks and Gelman's iterated graphical method to assess convergence \cite{brooks98}. This diagnostic method is based upon the covariance estimation of $J$ independent Markov chains of equal length $I$: $\{\Xv_{j,i};j = 1,\cdots,J;i = 1,\cdots,I\}$.
The intra-chain and inter-chain variances that characterize the convergence behaviour, are defined as covariance matrix averages:
\begin{align*}
\Vv_\text{intra} &= \fracp{1}{J(I-1)}\sum_j \sum_i \left(\Xv_{j,i} -  \overline{\Xv}_{j} \right) \left(\Xv_{j,i} -  \overline{\Xv}_{j} \right)^T \\
\Vv_\text{inter} &= \fracp{1}{J-1}\sum_j \left(\overline{\Xv}_{j} -  \overline{\Xv} \right) \left(\overline{\Xv}_{j} -  \overline{\Xv} \right)^T
\end{align*}
where $\overline{\Xv}_{j}$ and $\overline{\Xv}$ denote the local and the global mean of the chains respectively.
Brooks and Gelman \cite{brooks98} proposed to evaluate the multivariate potential scale reduction factor (MPSRF) given by
$$
R = \frac{I-1}{I}+\frac{J+1}{J} \lambda\pth{\Vv_\text{intra}^{-1}\Vv_\text{inter}}
$$
where $\lambda(\Av)$ denotes the largest eigenvalue of the matrix $\Av$.
Convergence is diagnosed when the MPSRF is close to one. As suggested in \cite{brooks98}, we have chosen $R < 1.2$.

\begin{figure}[htbp]
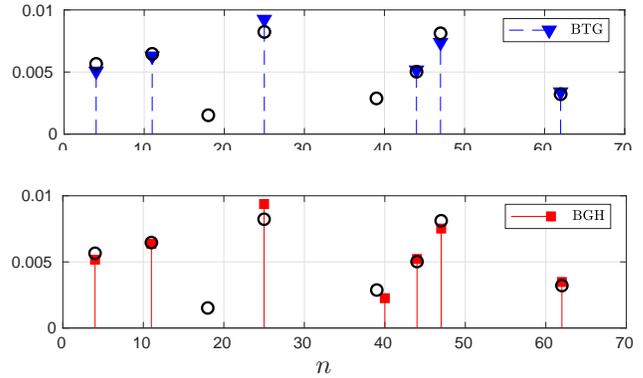

\begin{tabular}{@{}l@{~}c@{}}
\rotatebox{90}{\hspace{.9cm}$x_n$}
&\includegraphics[width=.45\textwidth]{./figs/./SSP_X_BGP_QHXLV_X} \\[-.1cm]
&$n$\\
\rotatebox{90}{\hspace{1cm}$x_n$}
&\includegraphics[width=.45\textwidth]{./figs/./SSP_X_BGH_RJ_Chol_QWHXLV_X}\\[-.1cm]
&$n$\\[-.2cm]
\end{tabular}
\caption{Estimation results at convergence of ${\xb}$ for the BTG (top figure) and BGH (bottom figure) compared to its true value (black circles).}
\label{fig:Estimation}
\end{figure}

Fig.~\ref{fig:Estimation} shows the Posterior Mean (PM) estimates of $\xb$ produced by both samplers.  They are of similar quality, which indicates that the approximation method of the truncated Gaussian by the GH distribution is effective.

\begin{figure}[htbp]
\centering
\includegraphics[width=.8\columnwidth]{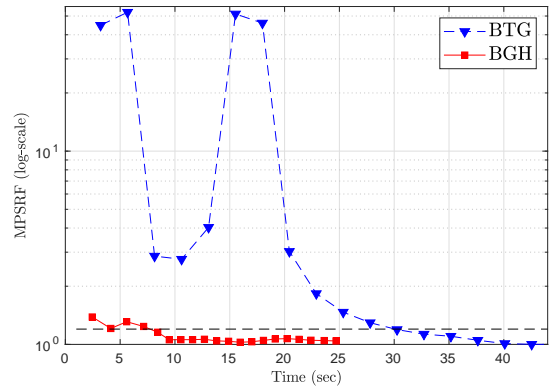}
\caption{Evolution of MPSRF (in log scale) of the BTG and BGH samplers. The horizontal line corresponds to the 1.2 threshold.}
\label{fig:MPSRF}
\end{figure}

Fig.~\ref{fig:MPSRF} gives the MPSRF of the variables $\qb$ in logarithmic scale for both samplers, as a function of time. Simulations were run using MATLAB on a computer with Intel Xeon E5-2680 processors with a CPUs clocked at 2.8 GHz.
To evaluate the MPSRF, each chain is divided into batches of $b$ samples, and the MPSRF is calculated upon the second halves of the Markov chains $\{\Xv_{j,i}\}, i=1,\cdots,kb$ for increasing lengths $kb$.
As recommended in \cite{brooks98}, we have chosen $b \approx \frac{I}{20}$, where $I$ is the length of the Markov chain. 
Whereas the BTG sampler needs around 36000 iterations to converge, the proposed BGH converges in about 1250 iterations, thanks to partially collapsed sampling. In terms of computing time, the acceleration factor is still significant but more modest (8 versus 30 seconds), since BGH iterations are more complex than that of BTG, despite the fact that the former have been optimized using recursive Cholesky factor updatings. Finally, let us stress that the considered numerical test corresponds to a problem of moderate complexity. For problems of increasing complexity, the relative efficiency of the partially collapsed sampler is expected to increase compared to the standard sampler, according to \cite{ge11}.
    
\section{Conclusion}
\label{sc:conclusion}
We have presented a new probabilistic hierarchical model for solving non-negative sparse signal restoration problems in an unsupervised way using MCMC methods. Thanks to the properties of GH distributions in terms of Gaussian mixture decomposition, the proposed model allows us to marginalize the amplitudes and to devise a partially collapsed Gibbs sampler, with improved mixing properties compared to the standard Gibbs sampler.

An interesting perspective would be to extend this approach based on Gaussian mixture decomposition and partially collapsed sampling to other restoration problems incorporating more general constraints than non-negativity.

\bibliographystyle{abbrv}
\bibliography{These.bib,Extra.bib}

\end{document}